\documentclass[reprint, amsmath,amssymb,aps,prl]{revtex4-2}
\usepackage{CJK}

\usepackage{graphicx}
\usepackage{wrapfig}
\usepackage{dcolumn}
\usepackage{bm}
\usepackage{chemformula} 

\usepackage{xr-hyper}
\usepackage{hyperref}

\makeatletter

\newcommand*{\addFileDependency}[1]{
\typeout{(#1)}
\@addtofilelist{#1}
\IfFileExists{#1}{}{\typeout{No file #1.}}
}\makeatother

\newcommand*{\myexternaldocument}[1]{%
\externaldocument{#1}%
\addFileDependency{#1.tex}%
\addFileDependency{#1.aux}%
}
\myexternaldocument{SupportingInfo}

\begin{document}
\begin{CJK*}{GB}{}

\title{Femtosecond electronic and hydrogen structural dynamics in ammonia imaged with ultrafast electron diffraction}

\author{Elio G.~Champenois$^{1}$}
\author{Nanna H.~List$^{2}$}
\email{nalist@kth.se}
\author{Matthew Ware$^1$}
\author{Mathew Britton$^{1,3}$}
\author{Philip H. Bucksbaum$^{1,3}$}
\author{Xinxin Cheng$^4$}
\author{Martin Centurion$^5$}
\author{James P.~Cryan$^1$}
\author{Ruaridh Forbes$^4$}
\author{Ian Gabalski$^{1,6}$}
\author{Kareem Hegazy$^{1,3}$}
\author{Matthias C. Hoffmann$^4$}
\author{Andrew J. Howard$^{1,6}$}
\author{Fuhao Ji$^4$}
\author{Ming-Fu Lin$^4$}
\author{J. Pedro Nunes$^5$}
\author{Xiaozhe Shen$^4$}
\author{Jie Yang$^{4,7}$}
\author{Xijie Wang$^4$}
\author{Todd J.~Martinez$^{1,8}$}
\author{Thomas J.~A.~Wolf$^{1}$}
\email{thomas.wolf@stanford.edu}
\affiliation{$^1$Stanford PULSE Institute, SLAC National Accelerator Laboratory, Menlo Park, CA 94025, USA.}
\affiliation{$^2$Department of Chemistry, KTH Royal Institute of Technology, SE-10044 Stockholm, Sweden.}
\affiliation{$^3$Department of Physics, Stanford University, Stanford, CA 94305, USA.}
\affiliation{$^4$SLAC National Accelerator Laboratory, Menlo Park, CA 94025, USA.}
\affiliation{$^5$Department of Physics and Astronomy, University of Nebraska Lincoln, Lincoln, NE 68588, USA.}
\affiliation{$^6$Department of Applied Physics, Stanford University, Stanford, CA 94305, USA.}
\affiliation{$^7$Department of Chemistry, Tsinghua University, Beijing, China}
\affiliation{$^8$Department of Chemistry, Stanford University, Stanford, CA 94305, USA.}

\date{\today}

\begin{abstract}
Directly imaging structural dynamics involving hydrogen atoms by ultrafast diffraction methods is complicated by their low scattering cross-sections. Here we demonstrate that megaelectronvolt ultrafast electron diffraction is sufficiently sensitive to follow hydrogen dynamics in isolated molecules. In a study of the photodissociation of gas phase ammonia, we simultaneously observe signatures of the nuclear and corresponding electronic structure changes resulting from the dissociation dynamics in the time-dependent diffraction. Both assignments are confirmed by \textit{ab initio} simulations of the photochemical dynamics and the resulting diffraction observable. While the temporal resolution of the experiment is insufficient to resolve the dissociation in time, our results represent an important step towards the observation of proton dynamics in real space and time.


\end{abstract}

\maketitle

\end{CJK*}
Thermal and photochemical hydrogen and proton transfer reactions are among the most ubiquitous in chemistry and biology.\cite{reece_proton-coupled_2009,hammes-schiffer_theory_2010,goyal_tuning_2017}
Directly following hydrogen and proton dynamics with time-resolved experimental methods is complicated by their fast time scales down to the few femtosecond regime. Moreover, typical time-resolved spectroscopic techniques exhibit only indirect sensitivity to these dynamics, due to the preferential interaction of the light with the electrons rather than the nuclei. Direct sensitivity to the motion of the nuclei can in principle be achieved with novel time-resolved imaging methods, such as ultrafast X-ray\cite{minitti_imaging_2015,stankus_ultrafast_2019,ruddock_deep_2019} and electron diffraction.\cite{yang_imaging_2018,Yang2020_Pyridine,wolf_photochemical_2019,champenois_conformer-specific_2021,yang_direct_2021}

X-rays exclusively scatter off the electron density of a molecule (Thomson scattering).\cite{moreno_carrascosa_ab_2019} The information about relative positions of atoms in a molecule from X-ray diffraction largely originates from inner-shell electrons with strong localization and, thus, high density at the positions of the nuclei. Since hydrogens do not posses inner-shell electrons, the sensitivity of X-ray diffraction to hydrogens and hydrogen motion is extremely limited. In contrast, electrons scatter off the Coulomb potential of a molecule (Rutherford scattering), which contains contributions from both electrons and nuclei.\cite{centurion_ultrafast_2022} Therefore, the relative cross-section of hydrogen with respect to carbon is more than an order of magnitude higher for electron compared to X-ray diffraction.\cite{hubbell_atomic_1975} The observation of hydrogen motion has recently been demonstrated in an investigation of the energy dissipation from the O-H stretching mode of bulk water using megaelectronvolt ultrafast electron diffraction (MeV-UED).\cite{yang_direct_2021,nunes_liquid-phase_2020, lin_imaging_2021} Here, we demonstrate that MeV-UED can resolve the femtosecond excited-state hydrogen dynamics in dilute gas phase ammonia, photoexcited at $\sim$200~nm.

\begin{figure}[htp]
\includegraphics[width=1.0\columnwidth]{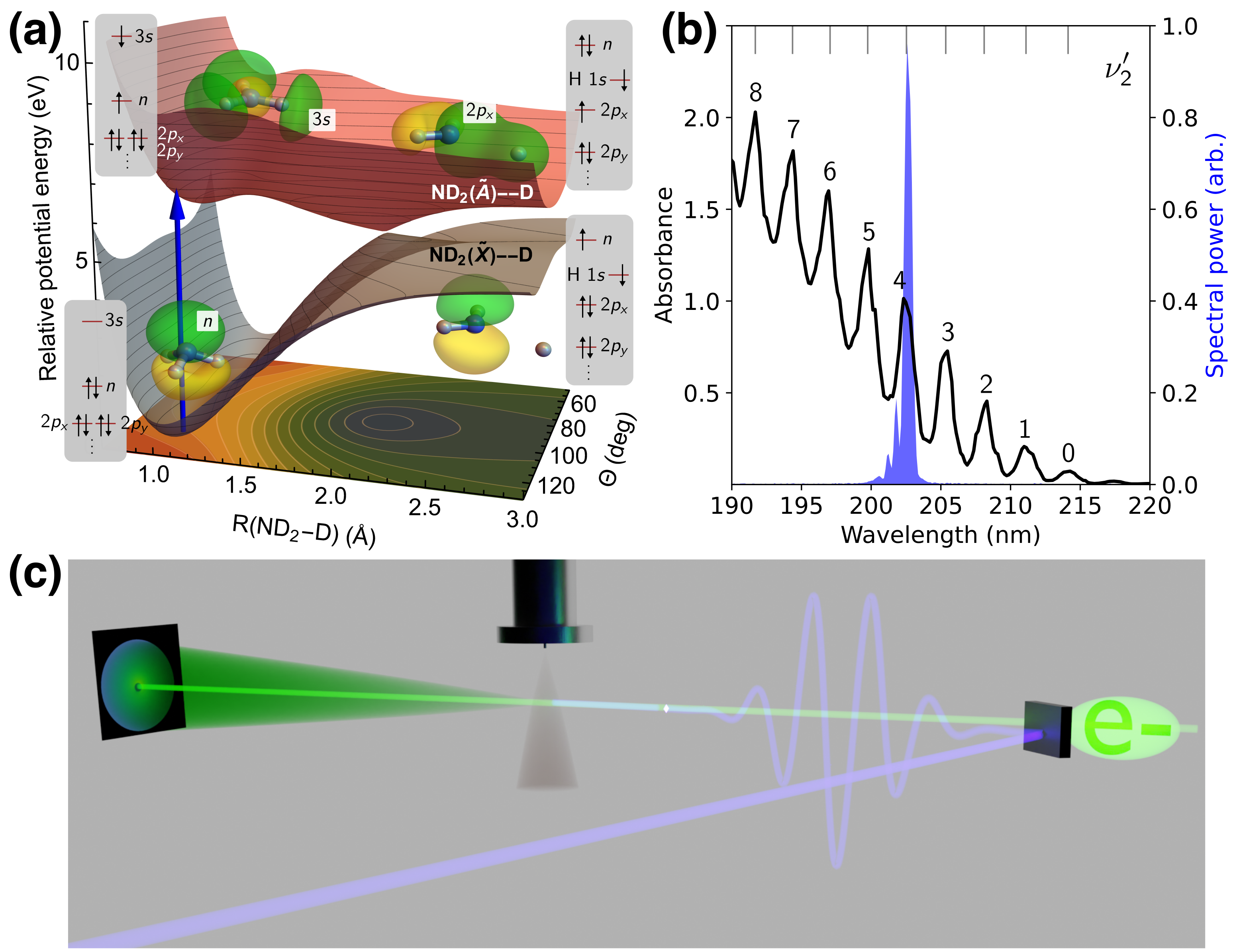}
\caption{\label{fig:epsart} Photoinduced process and experimental overview. (a) S$_0$ and S$_1$ PESs along the umbrella ($\Theta$) and \ch{ND2}-D dissociation coordinates. The PESs were obtained from Refs.~\citenum{Zhu2012_NH3pot,Jianyi2012_NH3photodiss}. Photoexcitation (blue arrow) to the predissociative 2$^1A$ state activates the umbrella mode and promotes adiabatic or non-adiabatic \ch{ND2}-D dissociation. The contour plot shows the S$_1$/S$_0$ energy gap, indicating a smaller gap (blue) along the \ch{ND2}-D coordinate. Changes in the electronic character are shown by means of the dominant configuration. Key orbitals involved in the process are shown as insets and correspond to state-averaged natural orbitals (isovalue=0.36~a.u.). At the distorted geometries, the Rydberg orbital correlates with a $\sigma^*$ orbital in the distorted \ch{ND2}-D and eventually becomes the $1s$ H orbital upon dissociation. Additionally, the dominant electron configurations (non-spin-adapted) of the states in the Franck--Condon region and in the dissociation limit are shown. (b) Experimental absorption spectrum of \ch{ND3} (black line) showing strong vibrational progression in the umbrella mode ($\nu_2'$). The peaks of the progression are labeled with respect to the corresponding vibrational level. The small peak at 217~nm results from a small contamination with \ch{NH3}. The spectrum of the UV pump laser centered around $\nu_2'=4$ of the umbrella mode is shown in blue. (c) Schematic diagram of the experimental setup. The UV pump (blue) and probe electron beam (green) are introduced to the interaction point at a 2$^{\circ}$ angle. Gaseous \ch{ND3} is introduced into the interaction point using a pulsed nozzle. Diffracted electrons are detected on a back-illuminated phosphor screen detector, while undiffracted electrons exit through a hole in the center of the detector.
}
\end{figure}

The photodissociation of ammonia is a benchmark case for multi-channel nonadiabatic photochemical dynamics and, therefore, has been the subject of many previous experimental steady-state\cite{vaida1987ultraviolet,ashfold_predissociation_1985,ashfold_resonance_1998,hause_vibrationally_2006,mordaunt_photodissociation_1996,rodriguez2014velocity} and time-resolved studies,\cite{chatterley2013,Evans2012,yu_tunneling_2014} as well as theoretical investigations.\cite{xie_communication_2015,Jianyi2012_NH3photodiss} Ammonia (C$_{3v}$ symmetry) exhibits a double minimum in its electronic ground state connected by an umbrella-type inversion motion (see potential energy surfaces, PESs, in Fig.~\ref{fig:epsart}a). Photoexcitation around 200~nm populates the $2^1A$ state, which is dominated by a single-electron excitation from the nitrogen lone pair ($n$) orbital to a $3s$ Rydberg orbital (see visualizations in Fig.~\ref{fig:epsart}a) and, therefore, exhibits Rydberg character. The $2^1A$ state has a minimum at a planar geometry of D$_{3h}$ symmetry. The large geometric difference between the ground state and $2^1A$ state results in a strong vibrational progression in its absorption spectrum (see Fig.~\ref{fig:epsart}b). Although \ch{H2N}-H dissociation is impeded by a barrier, the hydrogen atom can cross the barrier (potentially aided by tunnelling) leading to $<$100~fs lifetimes in the D$_{3h}$ minimum for all but the lowest two out-of-plane bending vibrational states. Isotopic substitution with deuterium significantly increases the lifetimes in the D$_{3h}$ minimum for a number of out-of-plane vibrational levels,\cite{vaida1987ultraviolet} making it easier to resolve the dynamics in time. Hence, we employ fully deuterated ammonia and excite the 4th excited state vibrational level at 202.5~nm (see Fig.~\ref{fig:epsart}b). Deuteration only affects the timescales here, and is not expected to have any effect on the diffraction signal intensity. Thus, our findings on the sensitivity of MeV-UED for deuterated ammonia are fully transferable to experimental signatures of proton dynamics in other systems.

The dissociation barrier results from the presence of an avoided crossing between the excited Rydberg state and a higher-lying $n\sigma^*$ state with \ch{D2N}-D antibonding character.\cite{Ashfold2010} Thus, the excited state gradually changes its electronic character from $n3s$ to $n\sigma^*$ along the \ch{D2N}-D dissociation coordinate. After passing the dissociation barrier, the wavepacket proceeds along the \ch{D2N}-D coordinate towards adiabatic and nonadiabatic photodissociation channels, yielding \ch{ND2}($\tilde{X}$)+\ch{D}
and \ch{ND2}($\tilde{A}$)+\ch{D} (see Fig.~\ref{fig:epsart}).

The sensitivity of electron scattering to the electronic structure of molecules has long been established and benchmarked, among other molecules, with the help of ammonia.\cite{bennani_differential_1979,duguet_high_1983,breitenstein_ci_1987} Elastic scattering processes are sensitive to the electron density in a molecule. In contrast, inelastic scattering is sensitive to electron correlation.\cite{bartell_effects_1964} We have recently demonstrated that inelastic scattering can be employed to follow electronic structure \textit{changes} during photochemical dynamics.\cite{Yang2020_Pyridine} Such changes in electron correlation can originate from population transfer between excited states of different electronic character through conical intersections\cite{Yang2020_Pyridine} or due to more gradual excited-state character changes like the change $n3s$ to $n\sigma^*$ when crossing the excited-state barrier of ammonia. The strongest contributions from inelastic scattering appear at momentum transfer values $<$2~$\mathrm{\AA^{-1}}$ whereas difference diffraction signatures of nuclear geometry changes can typically be measured up to 10~$\mathrm{\AA^{-1}}$.\cite{wolf_photochemical_2019} Thus, complementary information from electronic- and nuclear-structure changes can be detected in a well-separable fashion in the experimental diffraction patterns. In the present study, we observe, in addition to signatures of the structural N-D dissociation, clear signatures of the electronic excitation and electronic character change from $n3s$ to $n\sigma^*$.

The experiment was performed at the MeV-UED facility at SLAC National Accelerator Laboratory.\cite{shen_femtosecond_2019} Fig.~\ref{fig:epsart}c shows a schematic of the experimental setup. A 202.5~nm pump pulse was spatially and temporally overlapped with an electron pulse of 4.2~MeV kinetic energy in a pulsed jet of \ch{ND3}. Diffracted electrons were detected with a combination of a phosphor screen and a camera. The simulations were performed using \textit{ab initio} multiple spawning\cite{AIMS2000} (AIMS) in combination with \textit{ab initio} elastic and inelastic electron scattering simulations.\cite{Yang2020_Pyridine} A detailed description of both the experimental and theoretical methods can be found in Sec.~S1 of the supplemental material.

Our AIMS simulations are based on the PESs and nonadiabatic couplings reported by Yarkony \textit{et al.}\cite{Zhu2012_NH3pot,Jianyi2012_NH3photodiss} and provide a picture consistent with previous numerically exact quantum dynamics.\cite{Jianyi2012_NH3photodiss} In particular, following photoexcitation and progress along the \ch{D2N}-D dissociation coordinate (Sec.~S2 and Fig.~S3), 
the majority of the population undergoes nonadiabatic photodissociation ($\sim$66 \%) with a smaller portion ($\sim$24 \%) proceeding along the adiabatic channel (Fig.~S4). A small fraction ($<10$ \%) remains trapped (after 0.84~ps) on the excited state by the predissociation barrier and hence retains Rydberg character (Fig.~S5). 

\begin{figure*}[ht!]
\includegraphics[width=2.0\columnwidth,trim={3cm 1.2cm 2cm 1.cm},clip]{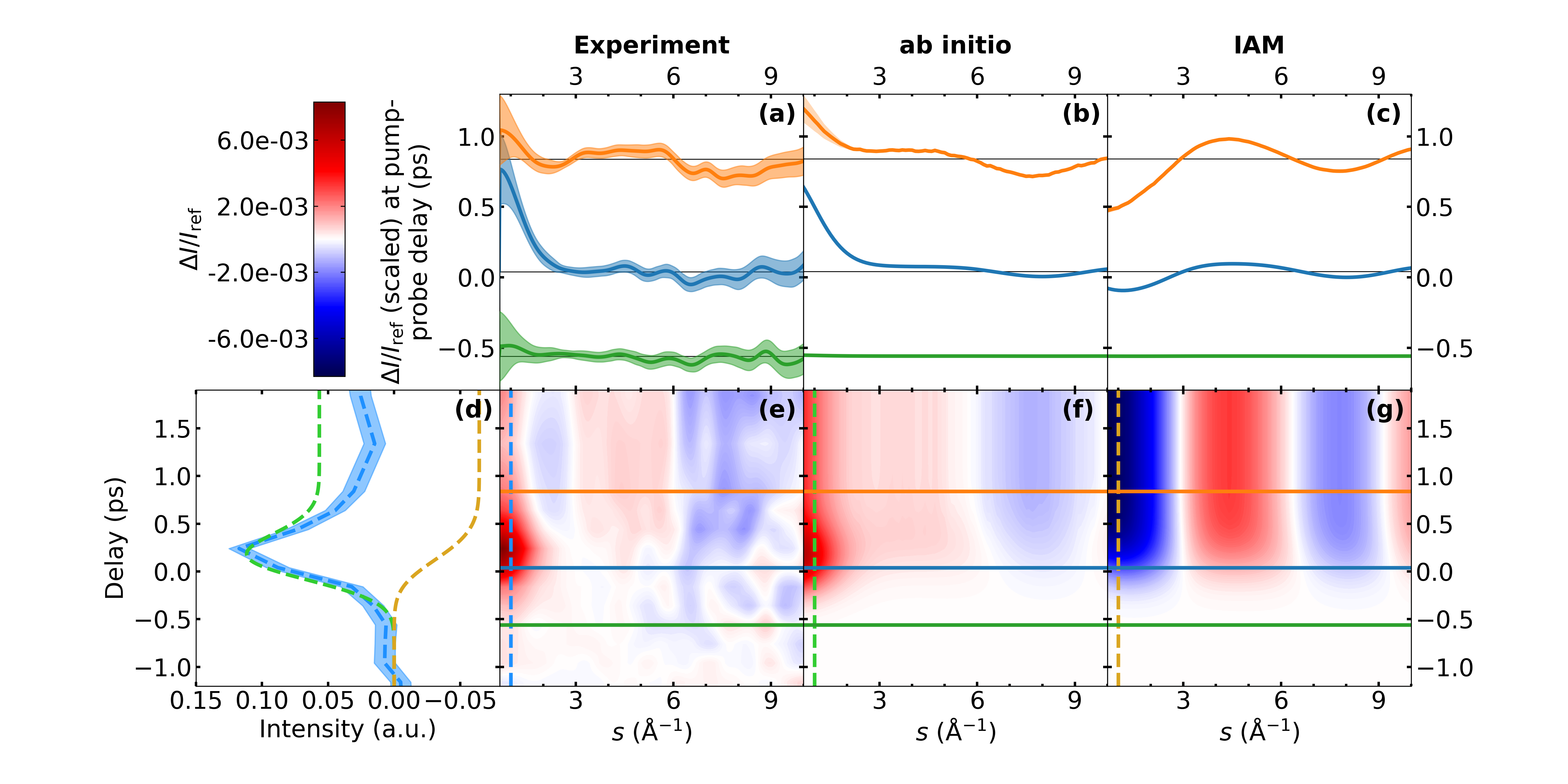}
\caption{\label{fig:contour} Comparison of experimental and simulated electron diffraction signatures of the photodissociation of ammonia. (a), (b), and (c) show $\Delta I/I_{\text{ref}}$ signals at different delays from (a) the experiment, (b) \textit{ab initio} scattering calculations based on AIMS simulations, and (c) the same AIMS simulations, but using the independent atom model (IAM) to compute the diffraction signal. The delays for the temporal lineouts in (a)-(c) are marked as color-coded horizontal lines in the corresponding false-color plots of the time-dependent signals from the experiment and the two different simulation approaches in (e)-(g). Additionally, the time-dependence of the integrated low-$s$ regions of the three false-color plots is shown in plot (d) where the upper-$s$ integration limits are marked by vertical color-coded dashed lines in plots (e)-(g).}
\end{figure*}

In Fig.~\ref{fig:contour}e, we present the results of our MeV-UED experiment as a false-color plot in the form of the difference between the time-dependent diffraction and the static diffraction of \ch{ND3} normalized by the static diffraction ($\Delta I/I_{\text{ref}}(s,t)$). Additionally, Fig.~\ref{fig:contour}a shows $\Delta I/I_{\text{ref}}(s)$ at the three different delay times ($t =$ -0.56~ps, 0.04~ps, and 0.84~ps), which are marked by the color-coded horizontal lines in Fig.~\ref{fig:contour}e. These delay times are chosen to include one delay clearly before time zero as a reference for the noise level of the experimental signals, the closest experimental delay to time zero, and one delay where the dissociation reaction is expected to be finished. The $s$-integrated \mbox{$\Delta I/I_{\text{ref}}(0.67 \ \mathrm{\AA}^{-1}<s<1 \ \mathrm{\AA}^{-1},t)$} (dashed blue vertical line in Fig.~\ref{fig:contour}e)  is shown in Fig.~\ref{fig:contour}d, blue. At time zero, a strong positive feature turns on in the $s<2$ $\mathrm{\AA}^{-1}$ regime and decays within the instrument response function (500-fs full width at half maximum, FWHM) to a weaker, delay-independent level (Fig.~\ref{fig:contour}d). Simultaneously, substantially weaker features appear: specifically, a broad positive signature between 3 and 6~$\mathrm{\AA}^{-1}$ and a broad negative signature at $s>6$ $\mathrm{\AA}^{-1}$ that stay constant over the whole remaining time delay window.

The experimental results in Fig.~\ref{fig:contour}e are compared with the simulated scattering signals computed based on AIMS dynamics of photoexcited \ch{ND3} (see Secs.~S1 D and S2). We use two different approaches to simulate the electron diffraction observable from the simulated wavepacket. First, using \textit{ab initio} scattering where the $\Delta I/I_{\text{ref}}(s,t)$ signatures are computed by scattering off the Coulomb potential from the nuclei and the electronic wavefunction as evaluated in our wavepacket simulations (Figs. \ref{fig:contour}b/f). This scattering simulation includes both elastic and inelastic scattering contributions. Analogous to the experimental data, the time-dependence of the integrated difference diffraction $0.67 \ \mathrm{\AA}^{-1}<s<1 \ \mathrm{\AA}^{-1}$ is plotted in Fig.~\ref{fig:contour}d (dashed green line). Second, we also provide $\Delta I/I_{\text{ref}}(s,t)$ signatures based on the independent atom model (IAM, Figs.~\ref{fig:contour}c/g) that neglect both inelastic scattering and changes in electron density around individual atoms due to chemical bonding or electron density redistribution (e.g., following electronic excitation). The time-dependence of the integrated signal at low $s$-values is plotted in Fig.~\ref{fig:contour}d (dashed brown line).

To understand the time-dependent signatures in Fig.~\ref{fig:contour}, we begin by considering the respective signatures at late delays (Fig.~\ref{fig:contour}a-c, orange). The signature from the IAM scattering simulation in Fig.~\ref{fig:contour}c exclusively originates from changes in the nuclear geometry (interatomic distances) due to the atomic superposition approximation inherent to IAM. Since it is a $\Delta I/I_{\text{ref}}(s,t)$ signature, it results from the difference of signatures from the atomic distances in and between the dissociation products at a delay time of 0.84~ps and signatures from the distances in the \ch{ND3} reactant geometry. Due to the relatively large atomic form factor of nitrogen, it is dominated by the difference of N-D distance signatures from the evolving photoexcited population and from the reactant geometry (see Sec.~S3 for details). Moreover, the presence of the predissociation barrier leads to a blurring of the diffraction signatures from the photoexcited population. This effect is further increased by the limited time resolution of the experiment (modeled in the IAM simulations in Fig.~\ref{fig:contour} by convolution with a Gaussian in time, see Sec.~S1 D). The signature at late delays in the IAM simulation (Fig.~\ref{fig:contour}c) is, therefore, dominated by the loss of one N-D bond distance (see Sec.~S3).

\begin{figure}[ht!]
	\centering
	\includegraphics[width=0.8\columnwidth]{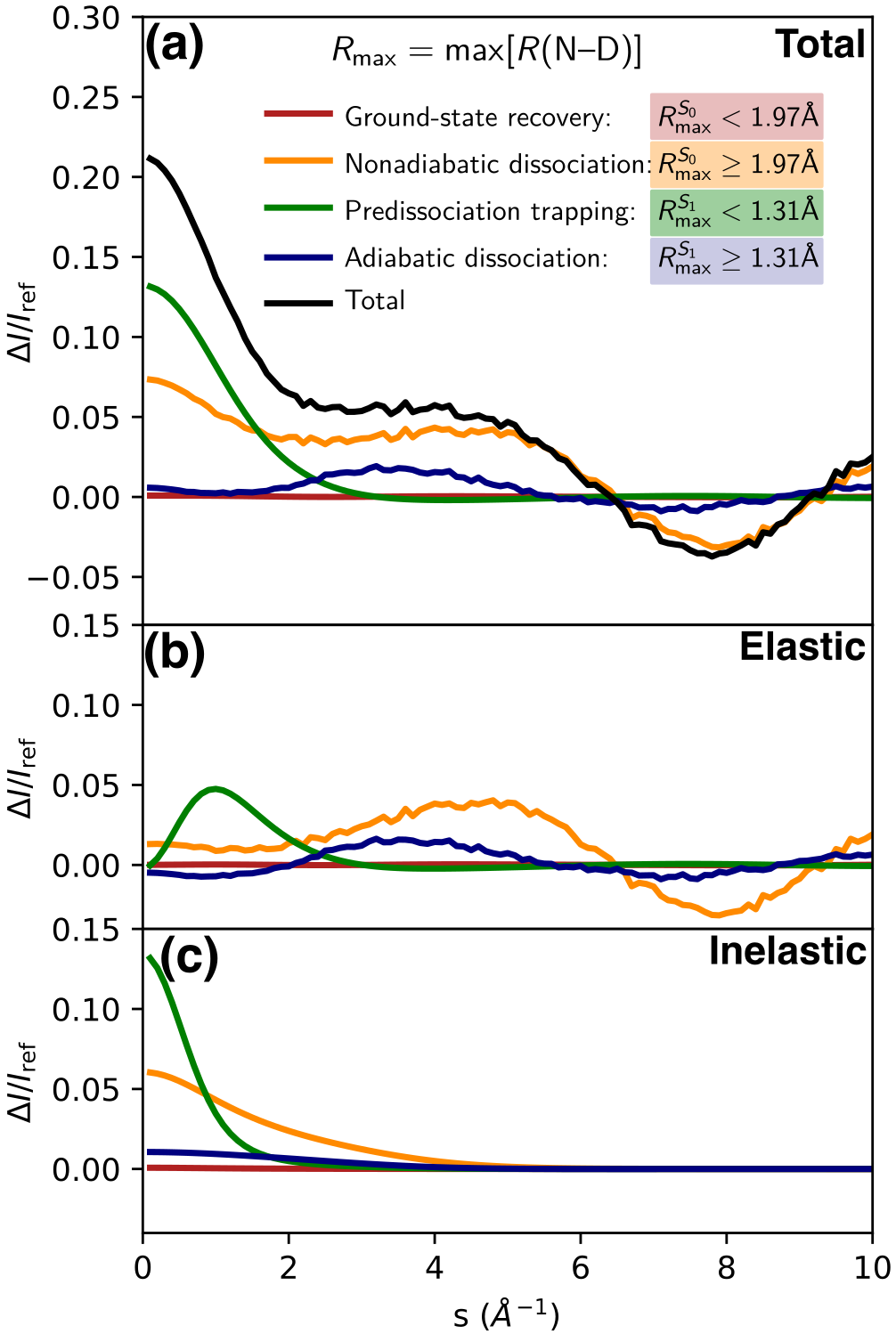}
	\caption{Decomposition of the longer timescale \textit{ab initio} scattering signal ($t=0.84$~ps) according to exit channel. (a) Total scattering signal, and its decomposition into (b) elastic and (c) inelastic components. The main contribution to the low-$s$ signal is the fraction of S$_1$ population trapped by the predissociation barrier. The distance-based cutoffs used to define the photoproducts are indicated in the figure (see also Fig~S4).
	}
	\label{fig:decomposition}%
\end{figure}

A qualitatively similar, but weaker, signal is found both in the experimental and \textit{ab initio} scattering signatures for $s>3$ $\mathrm\AA^{-1}$ (Figs.~\ref{fig:contour}a and b). A decomposition of the \textit{ab initio} scattering signature into elastic and inelastic scattering contributions (Figs.~\ref{fig:decomposition} and S6) shows that it is exclusively due to elastic scattering. Due to the incoherent nature of inelastic scattering, we do not expect any direct signatures from changes in the molecular geometry.\cite{Yang2020_Pyridine} Thus, the absence of the signature at $s>3$ $\mathrm\AA^{-1}$ in the inelastic scattering supports the assignment of this signature as the loss of one N-D bond, analogous to the IAM scattering signature. The relative weakness of the signatures with respect to the IAM simulation is a direct result of the deviation of the actual electron distribution in \ch{ND3} from a superposition of atomic densities assumed in IAM.

Considering next the low-$s$ ($s<3$~$\mathrm\AA^{-1}$) region at late delays (orange plots), the \textit{ab initio} scattering and experimental signals deviate qualitatively from the IAM. Both show a strong positive signature for $s<1$~$\mathrm\AA^{-1}$ whereas the IAM simulations exhibit a negative signature. Decomposition of the \textit{ab initio} scattering simulations (Figs.~\ref{fig:decomposition} and S6) show that its main contribution below 1~$\mathrm{\AA}^{-1}$ originates from the small fraction ($<10$ \%, Fig.~S4) of population that is trapped on the excited state due to the predissociation barrier. As shown in Fig.~S5, this is mainly due to the Rydberg character of the excited state rather than geometric changes from the difference between ground- state and excited-state potentials in the Franck--Condon region. 

It should be noted that both elastic and inelastic scattering contribute to the low-$s$ signature (see Figs.~\ref{fig:decomposition} and S6). Both contributions can be directly connected to the Rydberg character of the excited state. First, the strength of the elastic scattering contribution originates from the significant change in electron density distribution induced by the Rydberg excitation: one electron, i.e., \mbox{10 \%} of the overall electron density, is redistributed from a fairly localized lone pair orbital to a strongly delocalized Rydberg orbital. Second, the excitation from the electronic ground state (strongly correlated motion of the two electrons occupying the $n$ orbital) to a $n3s$ Rydberg state (weak correlation between remaining electron in the $n$ orbital and the electron in the $3s$ Rydberg orbital) can be expected to yield a significant change in the inelastic scattering signature. 

Weaker contributions to the $s<1$~$\mathrm\AA^{-1}$ signal arise from the large population ($\sim$66\%) in the nonadiabatic dissociation channel, mainly from inelastic scattering contributions (see Fig.~\ref{fig:decomposition}). The less populated adiabatic dissociation channel ($\sim$24\%) shows negligible contributions in this region. The disagreement between experimental and \textit{ab initio} scattering simulation between $1<s<2$ $\mathrm{\AA}^{-1}$ can be explained by a baseline offset in the experimental data in this region in combination with a slight overestimation of the inelastic scattering intensities in the \textit{ab initio} simulations.

Summing up our theoretical analysis at long time delays, we can distinguish two regions in the $\Delta I/I_{\text{ref}}(s,t)$ signal: (i) the $<1\ \mathrm{\AA}^{-1}$ region which is almost exclusively sensitive to the electronic character of the excited electronic state, and (ii) the $>3\  \mathrm{\AA}^{-1}$ region which is exclusive sensitivity to nuclear structural changes, i.e. the loss of an N-D bond upon photodissociation.

We can interpret the time-zero signal (blue curves in Fig.~\ref{fig:contour}a-c) along the same $s$-regions. Experimental and \textit{ab initio} scattering signals are dominated by a stronger signature at $s<1$ $\mathrm{\AA}^{-1}$ than at later delay times. The region at $s>3$ $\mathrm{\AA}^{-1}$ shows the same (albeit weaker) signatures as at later delay times. The strong signature at $s<1$ $\mathrm{\AA}^{-1}$ is easily explained by the fact that initially all of the excited-state population is residing behind the predissociation barrier, where the electronic state exhibits Rydberg character. Accordingly, the decay of the signal at $s<1$ $\mathrm{\AA}^{-1}$ after time zero is related to the depopulation of the quasi-bound Franck--Condon region due to the associated change in electronic character (see also Fig.~S5). Since IAM is insensitive to the electronic state, this feature is entirely missing in the IAM scattering signal. The weak dissociation signatures at $s>3$~$\mathrm{\AA}^{-1}$ are the result of a temporal smearing of the onset of the dissociation signatures due to the limited temporal resolution (see Fig.~S6 for the raw, i.e., without temporal convolution, theoretical difference scattering signals).

In conclusion, we observed signatures from both deuterium structural dynamics and electronic structure changes during the photodissociation of \ch{ND3} in well-separated momentum transfer ranges of ultrafast electron diffraction. Our results mark a powerful demonstration of the ability of MeV-UED to follow nonadiabatic proton and hydrogen photochemistry.
Alternative and emerging methods with similar or higher sensitivity to hydrogens and their dynamics are laser-induced electron diffraction\cite{pullen_imaging_2015,wolter_ultrafast_2016,sanchez_molecular_2021,belsa_laser-induced_2021} and Coulomb explosion imaging.\cite{boll_x-ray_2022,li_coulomb_2022} The sensitivity of these techniques to hydrogen has been established for static structures.\cite{boll_x-ray_2022,belsa_laser-induced_2021} Time-resolved studies of hydrogen motion with Coulomb explosion imaging have been recently demonstrated in a seminal study on the roaming reaction in formaldehyde.\cite{endo_capturing_2020} Structural dynamics studies with laser-induced electron diffraction have so far been limited to dynamics of the ionic states populated during its self-diffraction process with photoelectrons.\cite{belsa_laser-induced_2021} However, both methods lack the complementary electronic structure information which MeV-UED is able to deliver as a well-separable observable. Our results lack so far in temporal resolution (500~fs FWHM), a crucial parameter for the investigation of reaction dynamics involving hydrogens and protons. However, structural dynamics of more strongly scattering second row elements can already be investigated at the existing MeV-UED facility using lower electron pulse charges, which result in a 3-fold improved time resolution (150~fs FWHM).\cite{wolf_photochemical_2019}
The signal levels required for the observation of proton dynamics at this higher temporal resolution can be achieved, e.g., by an increase of the repetition rate. An increase of the repetition rate by several orders of magnitude has already been achieved with the electron injector guns for next-generation X-ray free electron lasers and, thus, does not pose a fundamental obstacle.

\begin{acknowledgments}
We thank Dr.~Xiaolei Zhu for helpful discussions. This work was supported by the AMOS program within the U.S. Department of Energy (DOE), Office of Science, Basic Energy Sciences, Chemical Sciences, Geosciences, and Biosciences Division. N.H.L.~acknowledges start-up funding from the School of Engineering Sciences in Chemistry, Biotechnology and Health (CBH), KTH Royal Institute of Technology. M.C.~ and J.P.F.N.~ acknowledge funding from the DOE Office of Basic Energy Sciences, Chemical Sciences, Geosciences, and Biosciences Division, AMOS program, under Award No. DE-SC0014170. The experiments were performed at the MeV-UED facility. MeV-UED is operated as part of the Linac Coherent Light Source at the SLAC National Accelerator Laboratory, supported in part by the U.S. Department of Energy (DOE) Office of Science, Office of Basic Energy Sciences, SUF Division Accelerator and Detector R\&D program, the LCLS Facility, and SLAC under contract Nos. DE-AC02-05CH11231 and DE-AC02-76SF00515. I.G. was supported by an NDSEG Fellowship.
\end{acknowledgments}

\bibliography{references}

\end{document}


\begin{CJK*}{GB}{}

\title[Supplemental Material:\\Femtosecond electronic and hydrogen structural dynamics in ammonia imaged with ultrafast electron diffraction ]{Supplemental Material for:\\ Femtosecond electronic and hydrogen structural dynamics in ammonia imaged with ultrafast electron diffraction}

\author{Elio G.~Champenois$^{1}$}
\author{Nanna H.~List$^{2}$}
\email{nalist@kth.se}
\author{Matthew Ware$^1$}
\author{Mathew Britton$^{1,3}$}
\author{Philip H. Bucksbaum$^{1,3}$}
\author{Xinxin Cheng$^4$}
\author{Martin Centurion$^5$}
\author{James P.~Cryan$^1$}
\author{Ruaridh Forbes$^4$}
\author{Ian Gabalski$^{1,6}$}
\author{Kareem Hegazy$^{1,3}$}
\author{Matthias C. Hoffmann$^4$}
\author{Andrew J. Howard$^{1,6}$}
\author{Fuhao Ji$^4$}
\author{Ming-Fu Lin$^4$}
\author{J. Pedro Nunes$^5$}
\author{Xiaozhe Shen$^4$}
\author{Jie Yang$^{4,7}$}
\author{Xijie Wang$^4$}
\author{Todd J.~Martinez$^{1,8}$}
\author{Thomas J.~A.~Wolf$^{1}$}
\email{thomas.wolf@stanford.edu}
\affiliation{$^1$Stanford PULSE Institute, SLAC National Accelerator Laboratory, Menlo Park, CA 94025, USA.}
\affiliation{$^2$Department of Chemistry, KTH Royal Institute of Technology, SE-10044 Stockholm, Sweden.}
\affiliation{$^3$Department of Physics, Stanford University, Stanford, CA 94305, USA.}
\affiliation{$^4$SLAC National Accelerator Laboratory, Menlo Park, CA 94025, USA.}
\affiliation{$^5$Department of Physics and Astronomy, University of Nebraska Lincoln, Lincoln, NE 68588, USA.}
\affiliation{$^6$Department of Applied Physics, Stanford University, Stanford, CA 94305, USA.}
\affiliation{$^7$Department of Chemistry, Tsinghua University, Beijing, China}
\affiliation{$^8$Department of Chemistry, Stanford University, Stanford, CA 94305, USA.}

\maketitle
\end{CJK*}

\newpage
\tableofcontents
\newpage
\section{Methods}\label{sec:methods}
\subsection{Experimental details}\label{sec:expdetails}
The experiment was performed at the MeV-UED facility at SLAC National Accelerator Laboratory.\cite{shen_femtosecond_2019} We used the 800 nm output of a Ti:Sapphire laser system and separated a pump and a probe beam path. The pulses of the probe beam path were frequency-tripled and directed onto the photocathode of an RF gun to eject an ultrashort electron pulse with a bunch charge of $\sim$8 fC at the interaction point after collimation. This charge is four times higher than reported in Ref.~\citenum{shen_femtosecond_2019} leading to a reduction of the time resolution to 500 fs full-width-half-maximum (FHWM) due to space-charge effects. The electrons were accelerated in a microwave cavity to a kinetic energy of 4.2 MeV and focused through a holey mirror to a spot size of 150 $\mu$m FWHM in the interaction region of a gas phase experimental chamber. Pump pulses were generated by frequency-quadrupling of the 800 nm laser pulses in a two-step process using $\beta$-barium borate crystals. The third harmonic, generated by an in-line trippler was mixed with the fundamental wavelength in a 100 $\mu$m thick $\beta$-barium borate crystal. The excitation wavelength of 202.5 nm was achieved by slightly detuning the angle of the mixing crystal. 
The resulting pulses of 15 $\mu$J/pulse were focused into the experimental chamber to a diameter of 280 $\mu$m FWHM and overlapped with the electron pulses at a 2$^{\circ}$ angle. Deuterated ammonia (99 atom \% deuteration) was purchased from Sigma-Aldrich and used without further purification. The isotope purity was monitored in situ with a residual gas analyzer. It was necessary to flow the sample through the apparatus for a few minutes to insure high isotope purity levels due to rapid proton/deuteron exchange with residual compounds in the experimental chamber (e.g. water). We used a pulsed nozzle (Parker, 100 $\mu$m orifice) in combination with a repetition rate of 180 Hz. Diffracted electrons were detected by a combination of a phosphor screen and an EMCCD camera. Based on the relative static and dynamic signal levels, the excitation ratio was estimated to be about 2.5 \%. Time-dependent diffraction was measured at a series of delay time points between $-3$ ps and +3 ps in each scan. The separation between time delay points was 200 fs, except for the earliest and latest delay points, where it was considerably larger. At each time delay point, we integrated the diffraction signal for 8 seconds. The full data set includes 400 such scans. The sequence of delay steps was randomized for every scan to avoid systematic errors.

\subsection{Generating $\Delta I/I_\text{ref}$ signals from the experimental raw images}
The data treatment was similar to those described in previous publications of results from the MeV-UED facility.\cite{champenois_conformer-specific_2021,wolf_photochemical_2019} The individual raw images from the Andor detector were treated in the following way: 
\begin{itemize}
    \item A corner of the image which does not contain contributions from the diffraction signal was used to subtract an intensity offset from the whole image.
    \item The detector exhibits a hole in the center to transmit the undiffracted electron beam. The hole was masked off in the individual images.
    \item The center of the diffraction pattern was determined in each image by filtering on pixels of a specific intensity and fitting a circle to the pixels. The obtained center coordinates were analyzed with respect to drift both in laboratory time and pump-probe delay time. We did observe slight drifts in laboratory time, but no drifts in delay time. The final center coordinates for each individual image were obtained from a smoothed curve describing the evolution of the center coordinates in laboratory time.
    \item The relativistic electrons create a sparse background of X-rays. X-rays which hit the detector typically saturate a number of adjacent pixels. Such X-ray hits were filtered out from each image by filtering on intensity values and identifying pixels, which deviated more than 3$\sigma$ from the mean of all pixel intensities with the same distance from the diffraction center.
    \item The total intensity of individual images was analyzed both for laboratory and pump-probe delay time drifts. Changes could only be observed on the laboratory time scale. They were corrected by normalization of each image to the mean intensity of all images of a specific pump-probe delay.
    \item Each image was azimuthally averaged to yield 1-dimensional diffraction patterns.
\end{itemize}

Diffraction patterns of the same pump-probe delay were averaged. The $\Delta I/I_\text{ref}$ signal was generated by subtracting the static diffraction signal from the diffraction of all time steps and dividing the resulting difference signal by the static diffraction signal. The $\Delta I/I_\text{ref}$ signal was Gaussian-smoothed in $s$ to remove high-frequency noise.

\subsection{Global fit of experimental $\Delta I/I_\text{ref}$ data}
\begin{figure}[ht!]
	\centering
	\includegraphics[width=1.0\columnwidth]{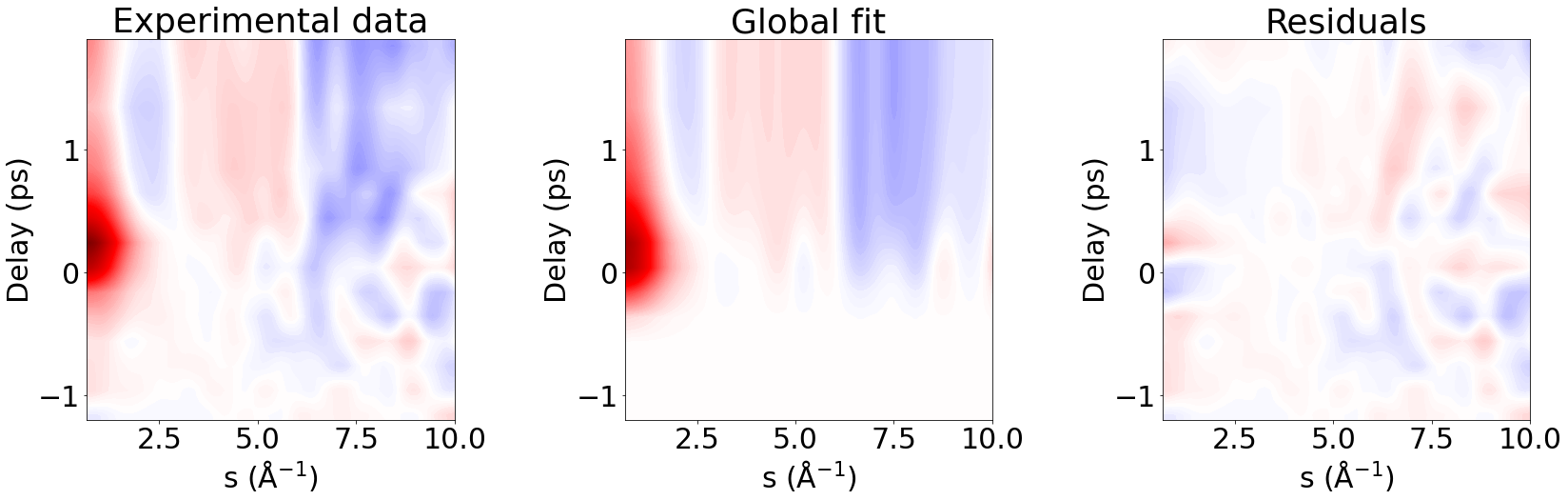}
	\caption{Comparison of a global fit of the $\Delta I/I_\text{ref}$ signal using Eq.~\ref{eq:fitfun} with the experimental data. The color scales for all three plots is identical with Fig.~2 in the main text. 
	}
	\label{fig:fit}%
\end{figure}

\begin{figure}[ht!]
	\centering
	\includegraphics[width=1.0\columnwidth]{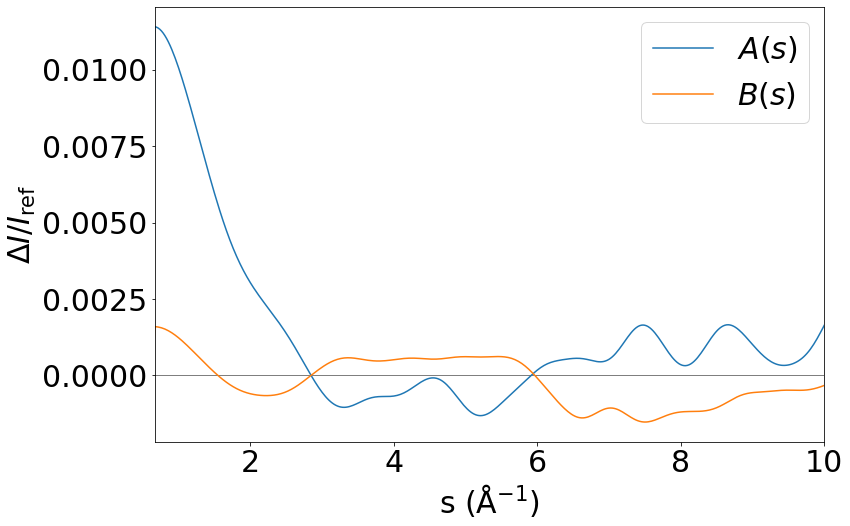}
	\caption{Decay-associated diffraction patterns $A\left(s\right)$ and $B\left(s\right)$ from fits using Eq.~\ref{eq:fitfun}. 
	}
	\label{fig:DAD}%
\end{figure}

The experimental $\Delta I/I_\text{ref}$ data were globally fitted with the following function:

\begin{equation}\label{eq:fitfun}
    \Delta I/I_\text{ref} \left(s, t\right) = g\left(\text{FWHM}, t\right) \ast \left[H\left(t\right)\cdot\left(A\left(s\right) e^{-\frac{t}{\tau}} + B\left(s\right) \right) \right].
\end{equation}
Eq.~\ref{eq:fitfun} represents a sum of an exponential and a constant term in the pump-probe delay $t$, multiplied by a Heaviside step function $H$. The product is convolved with a Gaussian function to account for the limited experimental time resolution. The FWHM of the Gaussian function was determined in the fit to be 0.46 ps, the time constant $\tau$ of the experimental decay to be 0.33 ps. Both values exhibit uncertainties on the order of the respective value. The global fit and its residuals are compared with the experimental data in Fig.~\ref{fig:fit}. The decay-associated diffraction patterns $A\left(s\right)$ and $B\left(s\right)$, which were determined in the fit (Eq.~\ref{eq:fitfun}), are plotted in Fig.~\ref{fig:DAD}.

\subsection{Computational details\label{sec:computdetails}}
\textit{Ab initio} multiple spawning\cite{AIMS2000} (AIMS) simulations were performed in the adiabatic basis using the 
potential energy surfaces for the $1,2{^1A}$ states of \ch{ND3}, their nuclear gradients and couplings reported by Yarkony and co-workers.\cite{Zhu2012_NH3pot,Jianyi2012_NH3photodiss}  
These were generated using multireference configuration interaction including single and double excitations (based on molecular orbitals from a two-state averaged complete active space expansion with eight electrons in nine orbitals) and the aug-cc-pVTZ basis set including a $3s$ Rydberg function on the nitrogen. A total of 48 initial conditions (ICs) were sampled from a 0K vibrational ground-state Harmonic Wigner distribution with  the requirement that the excitation energy is within $\pm0.015$ eV of the $v_2=4$ umbrella absorption maximum. 
Frequencies and normal modes were obtained using MP2 and the aug-cc-pVTZ basis set. 
This yields a total energy of $6.8{\pm}0.3$ eV, relative to the \ch{ND3}($\tilde{X}$) minimum, for the selected initial conditions. Each IC was initiated on S$_1$ under the independent first-generation approximation\cite{Hack2001} and propagated using AIMS for $\sim1$ ps ($4\cdot10^4$ a.u.) in a maximum time step of 10 a.u. ($\sim$0.24 fs). Since our implementation of \textit{ab initio} scattering is currently restricted to complete-active space self-consistent field (CASSCF) theory,\cite{Yang2020_Pyridine} we approximated the scattering signal on the basis of the CASSCF wavefunctions, accounting for potential state rotations through the zeroth-order XMS basis obtained at the extended multistate multireference second-order perturbation theory (XMS-CASPT2, real level shift of 0.3 a.u.). This allows us to retain the correct state-ordering from the dynamics but neglects the effects of first-order wavefunction corrections. However, no substantial rotations were observed in the XMS treatment, i.e., all rotation matrices were close to the identity matrix. The active space consisted of eight electrons in nine orbitals with state-averaging over the two lowest singlet states and the def2-svp basis with a $3s$ Rydberg function on the nitrogen (exponent 0.028 a.u.$^{-2}$). We computed the \textit{ab initio} scattering signals based on the AIMS dynamics in 2-fs time steps. Due to the large number of spawns, resulting in a large number of scattering calculations, we computed the two-dimensional diffraction pattern only in the $0<t<0.5$ ps window (see below). Additionally, we computed the difference scattering at 850 fs to get a late-time lineout. To further reduce the number of scattering calculations, we randomly selected five samples among the trajectory basis functions (TBFs) with population $<0.1$ for each IC and timestep, while retaining all TBFs with population $\ge0.1$. The wavepacket was renormalized accordingly.
Rotationally averaged elastic and inelastic scattering signals were evaluated using an $s$-grid ranging from 0.0 to 10.0 \AA$^{-1}$ (increments of 0.1 \AA$^{-1}$) and a 590-point Lebedev quadrature. The \textit{ab initio} scattering calculations were performed using the TeraChem program.\cite{TC1,TC2,TC3,Seritan21_TCreview}
For comparison, we also computed the scattering signal using the independent atom model (IAM). The atomic form factors were taken from Ref.~\cite{colliex_electron_2004}

Analogous to the experimental data, we generate a reference diffraction pattern (ref) from the ICs of the wavepacket simulations in the electronic ground state. We subtract this reference diffraction pattern from all simulated time steps and divide the resulting difference diffraction of each time step by it to obtain: $\Delta I/I_\text{ref}=\left(I_\text{total}(t)-I_\text{total}(t<0)\right)/I_\text{ref}(t<0)$ signal, where $I_\text{total}(t)=I_\text{elastic}(t)+I_\text{inelastic}(t)$. Decomposition of the total \textit{ab initio} signal into elastic and inelastic components was done as follows: $\left(I_\text{X}(t)-I_\text{X}(t<0)\right)/I_\text{total}(t<0)$ with $X=\text{elastic or inelastic}$, such that their sum yields $\Delta I/I_\text{ref}$.
The $\Delta I/I_\text{ref}$ signals of both the \textit{ab initio} scattering and the IAM scattering simulation are shown in Fig.~\ref{fig:2D_elastic_inelastic}. For a one-to-one comparison with the experimental diffraction signal (see Fig.~2 of the main text) and to account for the limited experimental time-resolution, we add negative time steps with $\Delta I/I_\text{ref} \left(t<0\right) = 0$. Furthermore, under the assumption that the changes in the $\Delta I/I_\text{ref}$ signal at delays beyond 500 fs are negligible, we used the last $\Delta I/I_\text{ref}$ pattern to extent the simulations out to 2 ps ($\Delta  I/I_\text{ref}\left(t>0.5 \mathrm{ps}\right)=\Delta I/I_\text{ref}\left(0.5 \mathrm{ps}\right)$). To account for the limited time resolution of the experiment, both the \textit{ab initio} and the IAM scattering simulations were convolved with a 500 fs FWHM Gaussian.

\section{Analysis of nonadiabatic dynamics simulations}\label{sec:SI_dynamics}
\begin{figure}[ht!]
	\centering
	\includegraphics[width=0.75\columnwidth]{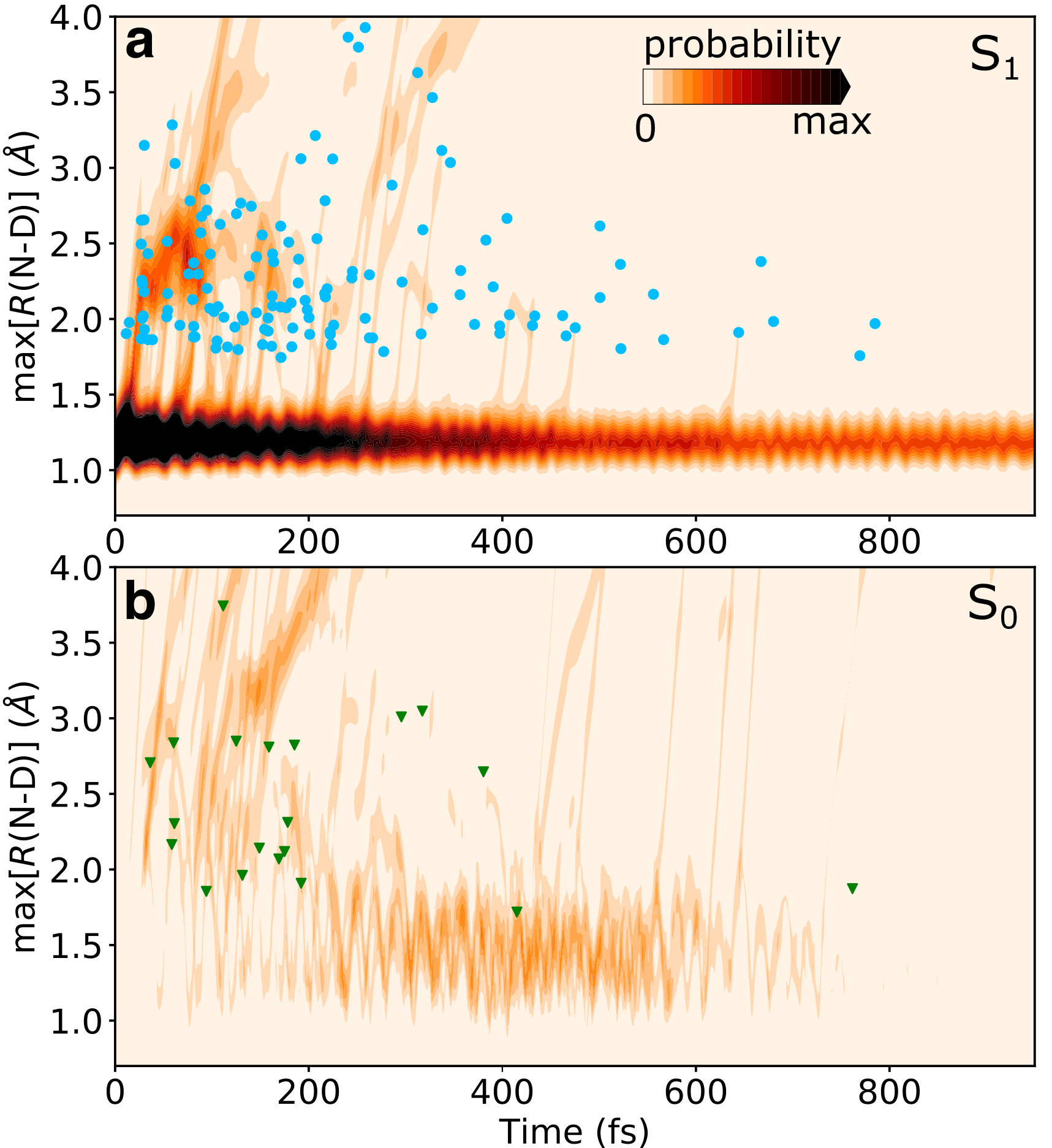}
	\caption{Time evolution of the wavepacket density projected onto the maximum N-D distance: (a) $\mathrm{S}_1$ and (b) $\mathrm{S}_0$. Blue filled circles and green triangles indicate the location of non-adiabatic $\mathrm{S}_1\rightarrow\mathrm{S}_0$ and $\mathrm{S}_0\rightarrow\mathrm{S}_1$ transition events, respectively. Additional transition events occurs at longer N-D distances (outside the plot range) due to Renner--Teller coupling. 
	}
	\label{fig:wavepacketprojection}%
\end{figure}

\begin{figure}[ht!]
	\centering
	\includegraphics[width=0.8\columnwidth]{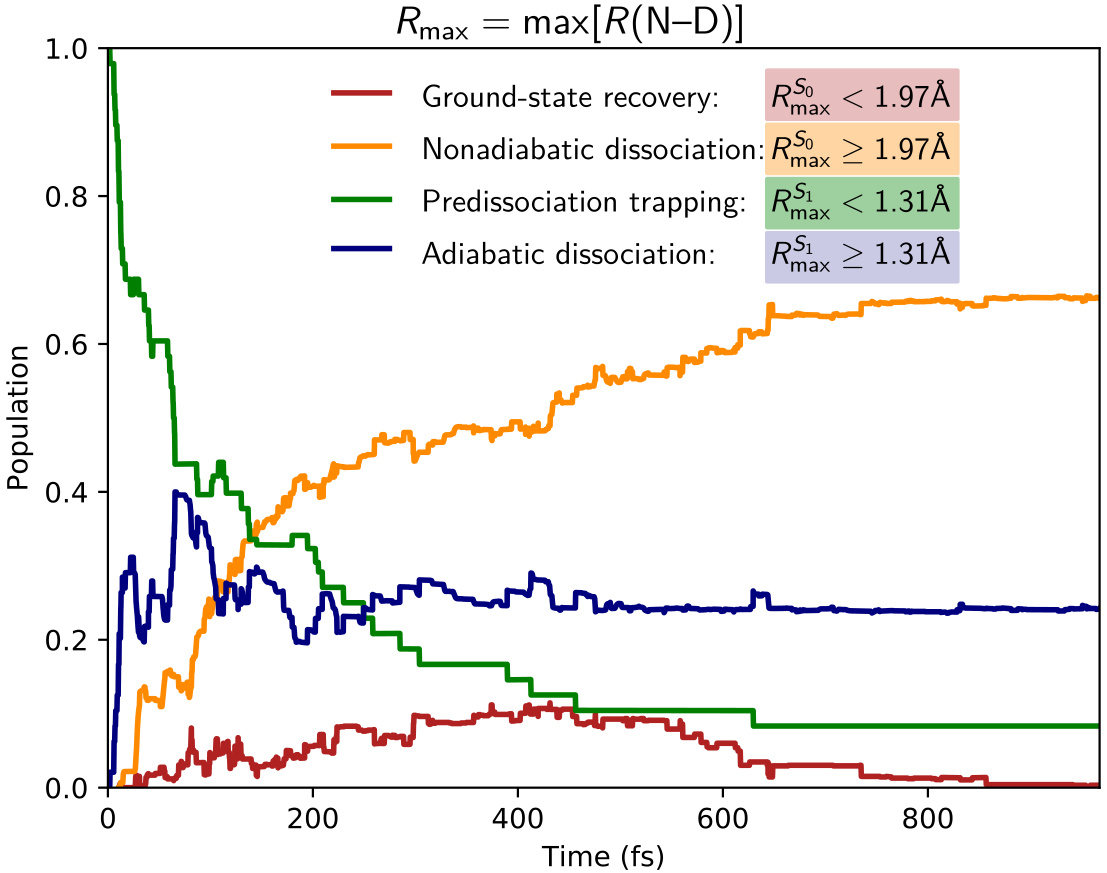}
	\caption{Population dynamics of the various photoproducts as obtained from the AIMS simulation.
	}
	\label{fig:pop}%
\end{figure}

In this section, we briefly describe the wavepacket behavior obtained from our AIMS simulations. Photoexcitation to the $n3s$ state initially triggers coherent oscillations along the umbrella mode with a period of ${\sim}50$ fs. The wavepacket proceeds along the N--D coordinate (see nuclear wavepacket densities in Fig.~\ref{fig:wavepacketprojection}) towards adiabatic and nonadiabatic photodissociation channels, yielding \ch{ND2}($\tilde{X}$)+\ch{D}
and \ch{ND2}($\tilde{A}$)+\ch{D}
photoproducts. In line with previous work,\cite{Dixon1988,Dixon1989} there are two types of exit channels depending on the recoil-induced momentum following deuterium photodissociation: (i) high rotational excitation of \ch{ND2} about its $a$-rotational axis caused by dissociation at out-of-plane configurations, and (ii) internal excitation of the bending mode in \ch{ND2}, leading to Renner--Teller coupling between the $\tilde{X}^2B_1$ and $\tilde{A}^2A_1$ states of \ch{ND2} (period of ${\sim}$30 fs).\cite{Rodriguez2014} Furthermore, the dissociation can take place adiabatically leading to \ch{ND2}($\tilde{X}$) state (Fig.~1 in the main text). Alternatively, the diabatic dissociation channel results in \ch{ND2}($\tilde{A}$) fragments. The temporal evolution of the product distribution is shown in Figure \ref{fig:pop}. As indicated in the figure, we used distance criteria based on the location of the predissociation barrier (${\sim}1.31$ \AA) and the minimum energy conical intersection (${\sim}1.97$ \AA) to separate the population according to exit channels.\cite{Jianyi2012_NH3photodiss} To distinguish adiabatic and nonadiabatic photodissociation, we further separate the population into contributions associated with S$_1$ and S$_0$. We find that about 8\% of the photoexcited population remains trapped by the dissociation barrier on S$_1$ after 1 ps. The population might be overestimated due to the classical evolution of the trajectory basis functions in AIMS and hence neglect of nuclear tunneling effects. Less than 1\% relaxes back to the \ch{ND3} minimum of the ground state. About 24 $\%$ of the photoexcited population undergoes adiabatic dissociation and ${\sim} 66$ \% nonadiabatic dissociation. The \ch{ND2}($\tilde{X}$,$\tilde{A}$) branching ratio agrees well with previous quantum dynamics simulations.\cite{Jianyi2012_NH3photodiss} 


\section{Signatures in the independent atom model simulations}\label{sec:IAM_signatures}
Within the independent atom model (IAM), the electron diffraction signature of an ensemble of randomly oriented molecules in momentum transfer space ($s$) can be expressed as
\begin{equation}\label{eq:IAM}
    I_{\text{tot}} = I_{\text{atom}} + \sum_{i=1}^{N} \sum_{j\neq i}^{N} f_{i}^{*}(s) f_j(s) \frac{\sin{r_{ij}s}}{r_{ij} s} = I_{\text{atom}} + I_{\text{mol}},
\end{equation}
where $I_{\text{atom}} = \sum_{i=1}^N \left|f_i(s) \right|^2$ is the atomic scattering intensity as the sum of the $N$ atomic contributions of the molecule, $f_{i}/f_{j}$ the atomic form factors of atoms $i$ and $j$, and $r_{ij}$ their distance. The double sum over all interatomic distances $r_{ij}$ in the molecule (molecular scattering, $I_{\text{mol}}$) contains the structural information of the molecule. The relative intensity of the individual $r_{ij}$ contributions to $I_{\text{mol}}$ is determined by the strengths of the atomic form factors. Due to the order-of-magnitude larger form factor of nitrogen with respect to deuterium, $I_{\text{mol}}$ is dominated by the signatures of the N-D distances while D-D distance have negligible contributions (see Fig.~\ref{fig:IAM_signatures}a). 

Within the IAM, individual atomic distances enter $I_{\text{mol}}$ as sinc functions with periods of $2 \pi / r_{ij}$ and zero crossings at $n\pi/r_{ij}$ for $n$ non-zero integers (see Eq.~\ref{eq:IAM} and Fig.~\ref{fig:IAM_signatures}). Moreover, only atomic distances which undergo changes enter difference diffraction ($\Delta I / I_{\text{ref}}(s, t)$), as a difference between the signature from the distance at time $t$ and the signature from the distance in the static (reference) diffraction. Figure \ref{fig:IAM_signatures}b, blue, shows a simulated $\Delta I / I_{\text{ref}}$ signature from elongating one of the N-D distances of \ch{ND3} by a factor of 10. For comparison, a $\Delta I / I_{\text{ref}}$ signature from elongating the atomic distance in a hypothetical molecule \ch{ND} is shown (orange), i.e., neglecting any D-D distance contributions to the signal. Analog to the static $I_{\text{mol}}$, the contributions from D-D distances are relatively small. In both curves, the signatures from the increased distance (oscillations with small period) and the reference distance (oscillations with large period, resembling the negative of the plots in Fig.~\ref{fig:IAM_signatures}a) can be clearly distinguished.

In the AIMS simulations, the predissociation barrier on the excited state leads to a blurred temporal onset of the photodissociation over a $\sim$100 fs range as shown in Fig.~\ref{fig:wavepacketprojection}. This results in a broad distribution of N-D distances between the photodissociation fragments at any given delay time. Moreover, due to the limited experimental time resolution, the measurement averages over a ~500 fs time window for every delay time and, thus, an even larger distribution of N-D distances. Therefore, the time-dependent N-D distances do not result in a distinct oscillatory contribution to the momentum transfer space $\Delta I/I_{\text{ref}}(s,t)$ signal based on the AIMS simulations. Accordingly, the $\Delta I/I_{\text{ref}}(s,t)$ signal is dominated at essentially all delay times by the loss of the well-defined initial N-D distance (green curve in Fig.~\ref{fig:IAM_signatures}).

\clearpage

\section{Supplemental figures}

\clearpage

\clearpage
\clearpage

\begin{figure}[ht!]
	\centering
	\includegraphics[width=0.7\columnwidth]{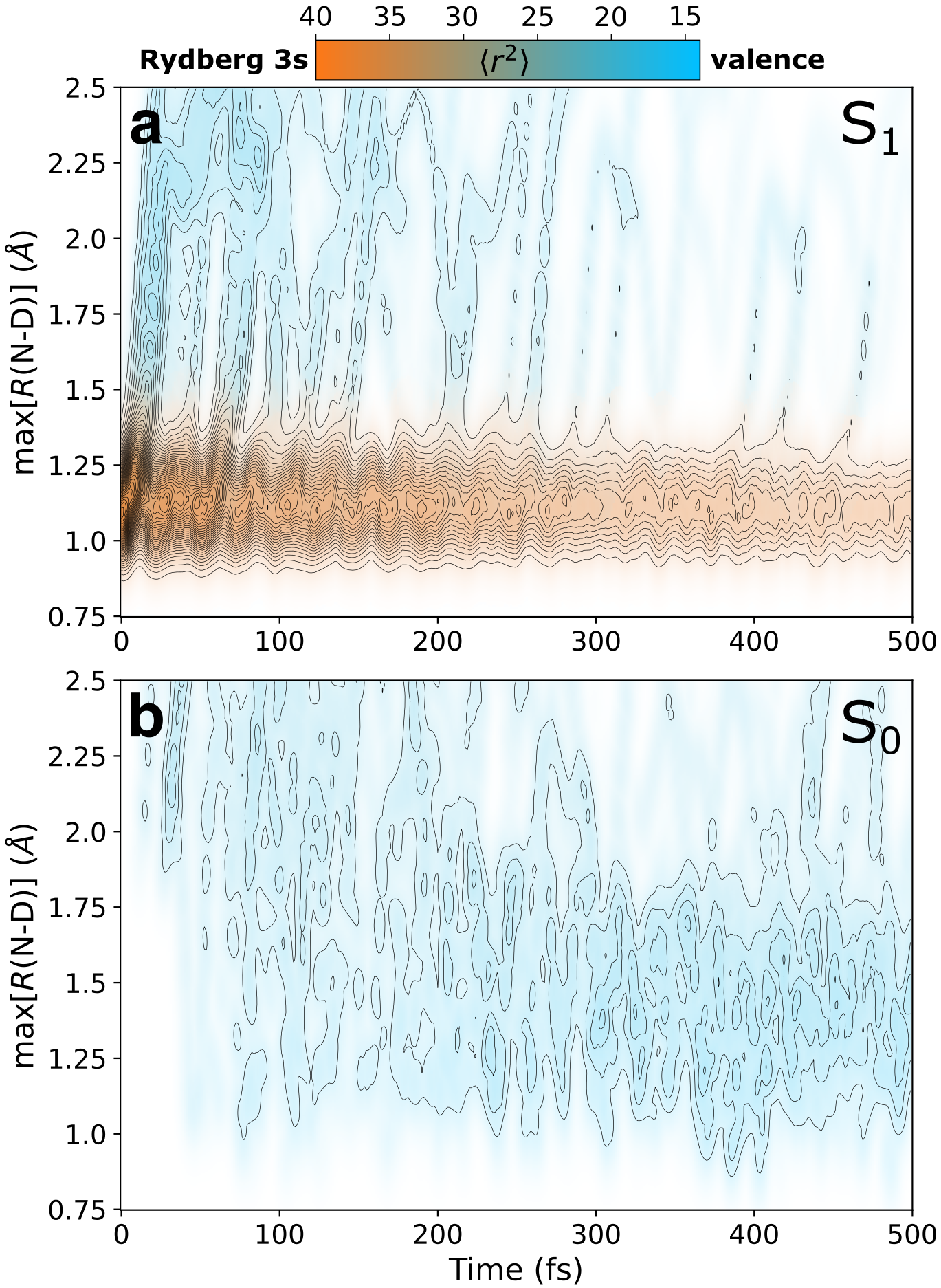}
	\caption{Time evolution of the nuclear wavepacket density along the N-D bond coordinate (zoom in of Fig~\ref{fig:wavepacketprojection}) for (a) $\mathrm{S}_1$ and (b) $\mathrm{S}_0$, with color-coding indicating the expectation value of $r^2$. The latter provides a measure for the radial extent of the electronic density, and hence the degree of Rydberg or valence character of the electronic state. Trapping by the predissociation barrier on S$_1$ is reflected in the substantial Rydberg character. At the Franck--Condon point, $\langle r^2\rangle_{\mathrm{S}_0}\sim 15$ a.u. and $\langle r^2\rangle_{\mathrm{S}_1}\sim 38$ a.u.
	}
	\label{fig:rydberg}%
\end{figure}

\clearpage

\begin{figure}[ht!]
	\centering
	\includegraphics[width=1.0\columnwidth]{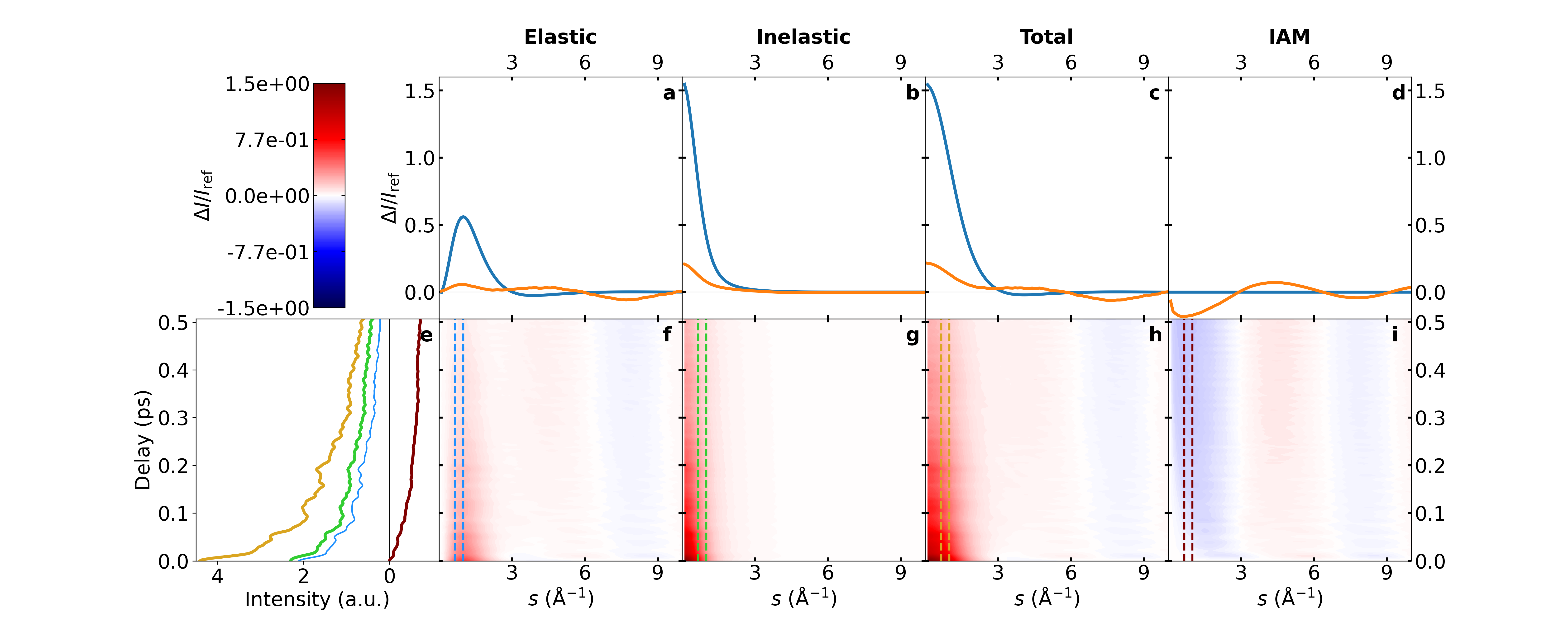}
	\caption{Comparison of the (a,f) elastic and (b,g) inelastic components to the total (c,h) \textit{ab initio} scattering and the independent atom model scattering (IAM, d,i). Lineouts of the diffraction pattern evolution are shown in plots a-d for 0 fs (blue) and 500 fs delay (orange). In the case of plots a-d, the blue curve represents signatures purely from the electronic structure change due to the photoexcitation, since the nuclei are in their initial positions. Since the IAM does not explicitly model the molecular electronic structure, the blue curve in plot d shows a zero line. Additionally, the time-dependence of the integrated regions, which are marked in (f-i) by dashed vertical lines, are shown in (e).
	}
	\label{fig:2D_elastic_inelastic}%
\end{figure}

\begin{figure}[ht!]
	\centering
	\includegraphics[width=1.0\columnwidth]{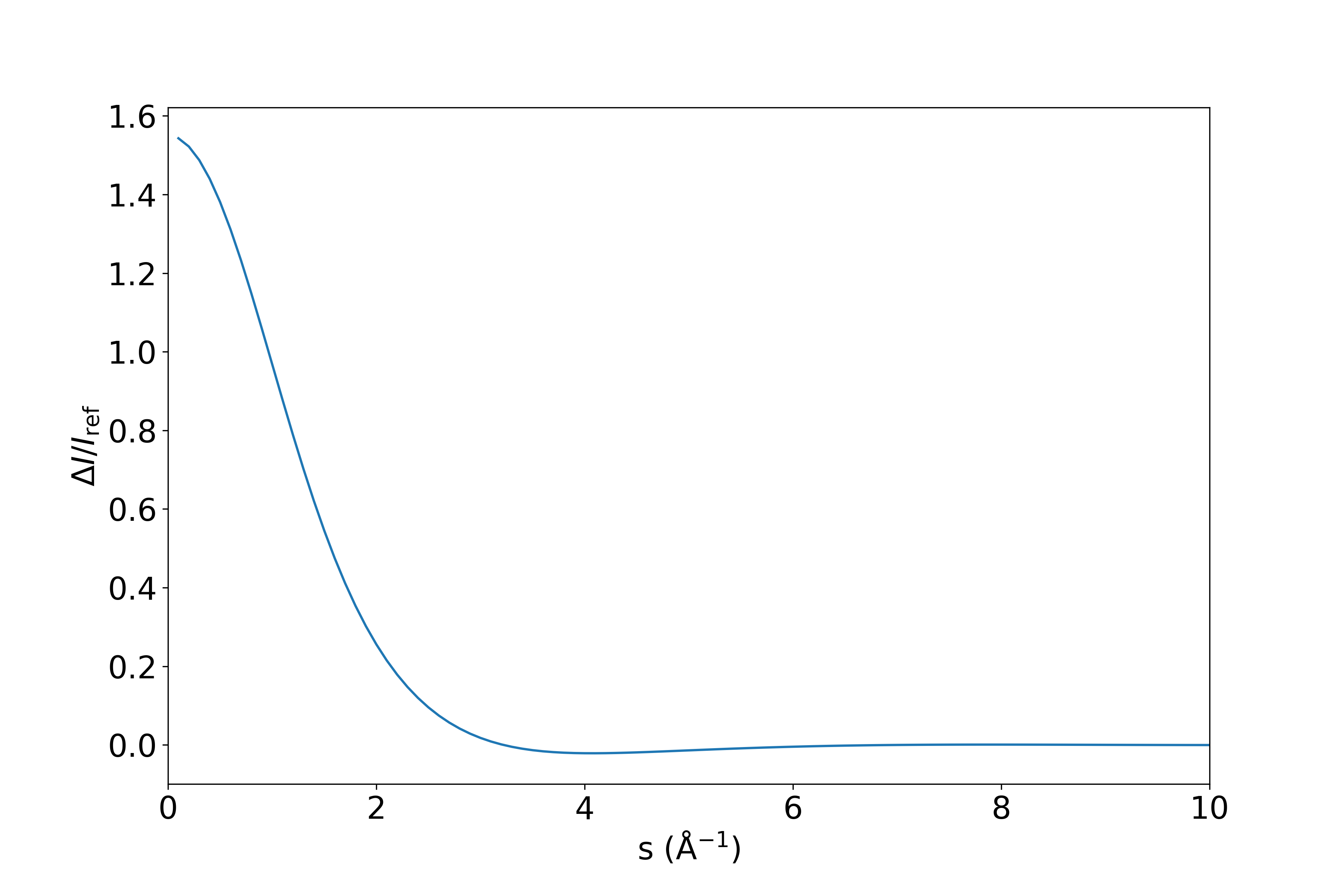}
	\caption{\textit{Ab initio} difference scattering signature of the electronic transition from the ground state to the $n3s$ state before the onset of any nuclear dynamics. In other words, the signatures are purely an effect of the changing electronic character upon photoexcitation.
	}
	\label{fig:FCpoint_signal}%
\end{figure}


\begin{figure}[ht!]
	\centering
	\includegraphics[width=1.0\columnwidth]{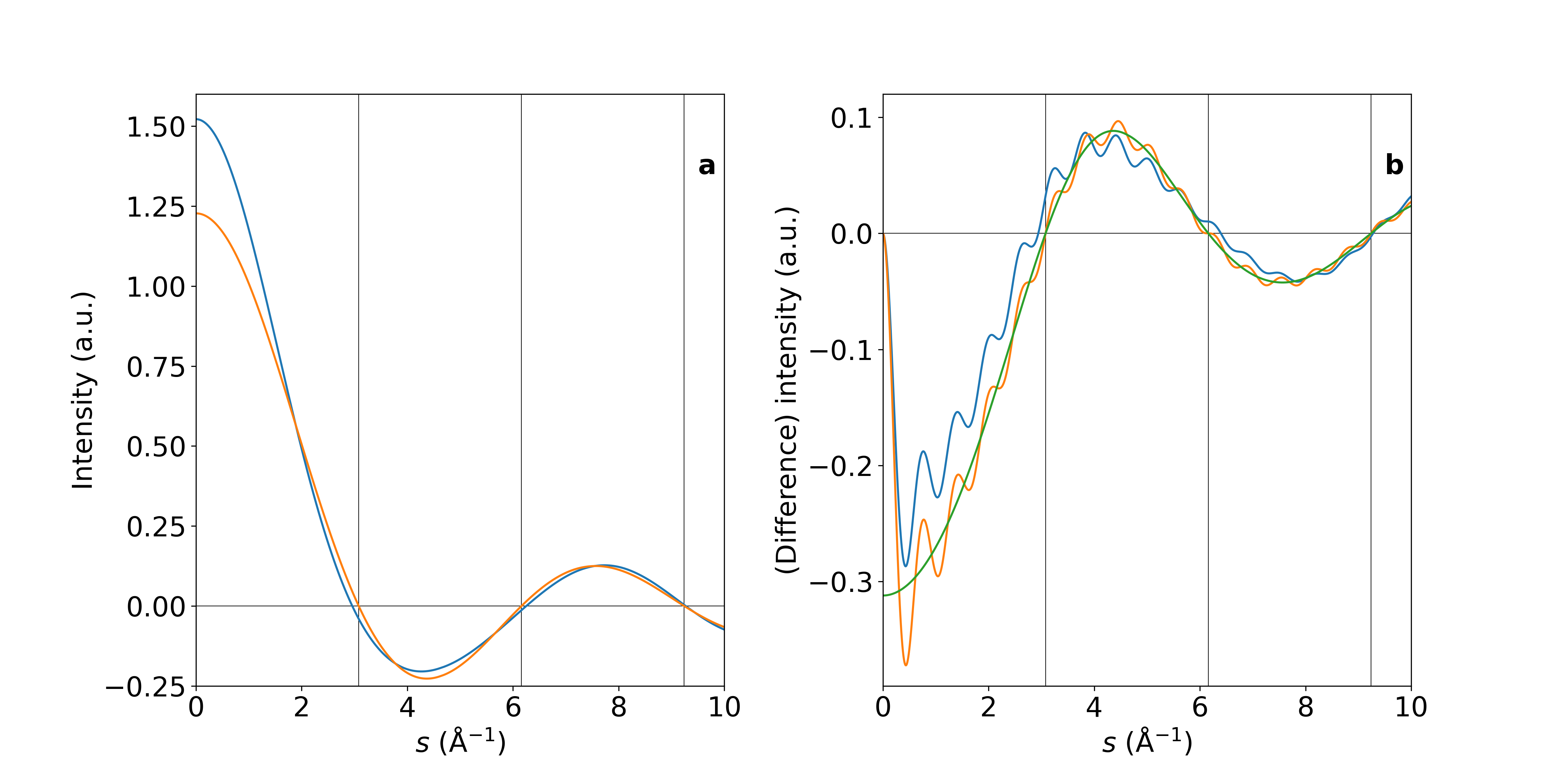}
	\caption{Zero-crossings in the independent atom model (IAM). (a) Simulated $I_{\text{mol}}$ signature of \ch{ND3} (blue) based on IAM. For better visibility of the behavior of the signature over the whole s-range, it is rescaled by $I_{\text{atom}}$. For comparison, the IAM signature of a single N-D bond distance is shown (orange). It is multiplied by a factor of three to account for the three N-D distances in \ch{ND3} and rescaled by $I_{\text{atom}}$. The two curves only deviate significantly for $s<2$ $\mathrm{\AA^{-1}}$ due to the contributions from D-D distances in \ch{ND3}. Their zero crossings at $\pi/r_{\mathrm{N-D}}$, $2\pi/r_{\mathrm{N-D}}$, and $3\pi/r_{\mathrm{N-D}}$ are marked. (b) Difference diffraction ($\Delta I / I_{\text{ref}}$) for the elongation of one of the N-D distances in \ch{ND3} by a factor of 10 (blue). For comparison, the $\Delta I / I_{\text{ref}}$ signature from the elongation of an isolated N-D distance (omitting contributions from the associated D-D distance changes) in \ch{ND3} is shown (orange). Additionally, the signature from the loss of a N-D distance is plotted in green. The fact that all curves share closely the zero crossings of part (a) demonstrates that the signals are dominated by the loss of an N-D bond distance.
	}
	\label{fig:IAM_signatures}%
\end{figure}

\clearpage

\bibliography{references}